\documentclass[conference]{IEEEtran}
\IEEEoverridecommandlockouts
\usepackage{cite}
\usepackage{amsmath,amssymb,amsfonts}
\usepackage{algorithmic}
\usepackage{graphicx}
\usepackage{textcomp}
\usepackage{multirow}
\usepackage{longtable}

\usepackage{xcolor}
\def\BibTeX{{\rm B\kern-.05em{\sc i\kern-.025em b}\kern-.08em
    T\kern-.1667em\lower.7ex\hbox{E}\kern-.125emX}}
\begin{document}

\title{Development of Space Qualified Signal Processing Readout Electronics for HabWorlds and Origins Space Telescope Detector Arrays
 \thanks{NASA Strategic Astrophysics Technology (SAT) 2025 Grant}
}

\author{\IEEEauthorblockN{1\textsuperscript{st} Tracee L. Jamison-Hooks}
\IEEEauthorblockA{\textit{Arizona State University School of Earth \& Space Exploration (SESE)} \\
\textit{Digital Signal Processing Technology Development Group}\\
Tempe, AZ, USA \\
Tracee.Jamison-Hooks@asu.edu}
\and
\IEEEauthorblockN{2\textsuperscript{nd} Lynn Miles}
\IEEEauthorblockA{\textit{NASA Goddard Space Flight Center (GSFC)} \\
\textit{Instrument Electronics Development Branch} \\
Greenbelt, MD, USA \\
}
\and
\IEEEauthorblockN{3\textsuperscript{rd} Sanetra Newman-Bailey}
\IEEEauthorblockA{\textit{NASA GSFC} Greenbelt, MD, USA \\
} 
\and
\IEEEauthorblockN{4\textsuperscript{th} Oketa Basha}
\IEEEauthorblockA{\textit{ASU Fulton Engineering and SESE} \\
Tempe, AZ, USA} 
\and
\IEEEauthorblockN{5\textsuperscript{th} Abarna Karthikeyan}
\IEEEauthorblockA{\textit{ASU SESE} Tempe, AZ, USA \\
}
\and
\IEEEauthorblockN{6\textsuperscript{th} Sarah E Kay}
\IEEEauthorblockA{\textit{Science Systems \& Applications Inc.} \\
Lanham, MD, USA \\
}
\and
\IEEEauthorblockN{7\textsuperscript{th} Sean Bryan
\IEEEauthorblockA{\textit{ASU SESE} Tempe, AZ, USA \\
}
\and
\IEEEauthorblockN{8\textsuperscript{th} Philip Mauskopf}
\IEEEauthorblockA{\textit{ASU SESE} Tempe, AZ, USA \\
}
}
\and
\IEEEauthorblockN{9\textsuperscript{th} Thomas Essinger-Hileman}
\IEEEauthorblockA{\textit{NASA GSFC Greenbelt, MD} \\
}
\and
\IEEEauthorblockN{10\textsuperscript{th}Jason Glenn}
\IEEEauthorblockA{\textit{NASA GSFC Greenbelt, MD} \\
}
\and
\IEEEauthorblockN{11\textsuperscript{th}Sumit Dahal}
\IEEEauthorblockA{\textit{NASA GSFC Greenbelt, MD} \\
\textit{Johns Hopkins University Baltimore, MD}\\ 
}
\and
\IEEEauthorblockN{12\textsuperscript{th}Adrian Sinclair}
\IEEEauthorblockA{\textit{NASA GSFC Greenbelt, MD} \\
}
\and
\IEEEauthorblockN{13\textsuperscript{th}Kathryn Chamberlin}
\IEEEauthorblockA{\textit{NASA JPL} Pasadena, CA \\
}
}

\maketitle

\begin{abstract}
The \textit{Habitable Worlds Observatory (HWO)}---a flagship ultraviolet/optical/infrared space telescope recommended by the \textit{National Academies’ Pathways to Discovery in Astronomy and Astrophysics}---will require detector technologies capable of supporting significantly larger pixel-count arrays than previous missions. \textit{Microwave Kinetic Inductance Detectors (MKIDs)}, naturally suited to microwave multiplexing readout, are already in use across several balloon-borne missions with FPGA-based systems. To transition this capability to space, we are developing a \textit{radiation-hardened detector readout system} that builds directly on the technical and environmental requirements defined by the \textit{PRIMA} mission. PRIMA serves as a critical pathfinder, informing the radiation tolerance, resource constraints, and on-board processing capabilities needed for HWO. In this work, we present our current results on algorithm implementation, hardware architecture, and firmware development using the radiation-hardened \textit{AMD Kintex Ultrascale FPGA}, aligning with PRIMA’s stringent specifications to ensure compatibility with future space-based observatories like HWO.

\end{abstract}


\section{Introduction}
\label{sect:intro}  
\begin{figure}[h]
\centering
\includegraphics[width=9cm]{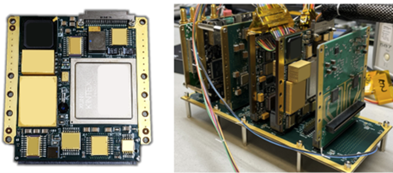}
\caption{The proposed readout system builds upon high-TRL modular SpaceCube hardware. Left panel: The readout system uses a build-to-print SpaceCube v3.0 Mini where the FPGA card (primary side pictured) houses a Xilinx Kintex KU060 chip. Right panel: Photograph of a full SpaceCube electronics system with 1U CubeSat cards inserted into a backplane.}
\label{figure:SpaceCube}
\end{figure}

The UV/optical/near-IR (UVOIR) is the recommended
waveband for NASA’s next flagship, the Habitable
Worlds Observatory (HWO; \cite{nasaHWO}). A key driving requirement
of HWO is to spectrally characterize earth-like planets,
a measurement which requires extremely low dark count rates.
Transition-edge-sensed (TES) bolometers and microwave kinetic
inductance detectors (MKIDs) are naturally excellent photon
counting detectors, with MKIDs in particular demonstrating $10^{-3}$
counts/sec/pixel \cite{hwoexoplanets}. MKIDs are superconducting resonators
that change their resonant frequency ($f_{0}$) and quality factor (Q) when incident radiation is absorbed in the superconducting material. Each MKID resonator can be designed to have its own unique starting resonant frequency, typically in the range of 0.05-5 GHz, and corresponding bandwidth. As a result, the resonator is said to be frequency-division multiplexed. Resonator array readout is accomplished by tracking each resonator frequency response in the transmission line and mapping amplitude, phase and frequency variation to the amount of kinetic inductance present and, hence, the properties of the incident photon. Our signal processing readout electronic methodology is designed to monitor the frequency and phase response of a large array of kinetic inductance detectors. Our firmware consists of a digital signal processing (DSP) chain that performs all needed readout functions for frequency-multiplexed detector arrays, which includes waveform generation, coarse and fine channelization (digital down conversion), time-stamping and data output via SpaceWire \cite{parkes2005spacewire} to an acquisition computer. The spaceflight readout electronics architecture builds upon high-heritage ($\ge$  TRL 6) SpaceCube hardware being adapted for the PRobe far-Infrared Mission for Astrophysics (PRIMA; \cite{prima}). SpaceCube, illustrated in Figure \ref{figure:SpaceCube}, is a family of NASA Goddard-developed space processors that established a hybrid-processing approach that leverages the advantages of commercial and radiation-hardened technologies \cite{nasaSpaceCube}.

\begin{table}[]
\centering
\caption{Driving requirements and architectural parameters for the MKID readout system derived from PRIMA mission specifications.}
\label{tab:PRIMAReq}
\small
\scalebox{0.7}
{
\begin{tabular}{|l|l|l|}
\hline
\textbf{Parameter} &
  \textbf{Value} &
  \textbf{Driver} \\ \hline
Number of Tones &
  1400 &
  1008 for science, 392 for blind/calibration \\ \hline
\begin{tabular}[c]{@{}l@{}}MKID frequency \\ placement\\ and recovery \\ precision\end{tabular} &
  10 kHz &
  $\frac{MKID_{BW}(\sim 30 \rm kHz)}{3}=10 ~\rm kHz$ \\ \hline
\multirow{3}{*}{\begin{tabular}[c]{@{}l@{}}Sampling \\ Frequency\end{tabular}} &
  \multirow{3}{*}{$\rm F_{s}=\rm 5 ~GHz$} &
  Science Bandwidth (BW) $\le$ 2.4 GHz \\ \cline{3-3} 
 &
   &
  ADC TI-ADC12J5200 \\ \cline{3-3} 
 &
   &
  DAC TI-DAC12DL3200 \\ \hline
\begin{tabular}[c]{@{}l@{}}Number of \\ Samples\end{tabular} &
  $2^{19}$ &
  $\frac{\rm F_{s}}{\rm 10kHz}=\frac{\rm 5GHz}{\rm 10kHz}$ \\ \hline
  
\multirow{2}{*}{\begin{tabular}[c]{@{}l@{}}Frequency \\ Translation\\ Architecture\end{tabular}} &
  \multirow{2}{*}{\begin{tabular}[c]{@{}l@{}}Coarse and Fine \\ Channelization\end{tabular}} &
  \begin{tabular}[c]{@{}l@{}}Coarse Channelization: 1024 Channel Oversampled \\ Polyphase Filterbank (PFB) \cite{harris2022multirate} \end{tabular} \\ \cline{3-3} 
 &
   &
  \begin{tabular}[c]{@{}l@{}}Fine Channelization: CORDIC-based Digital \\ Downconversation (DDC)\end{tabular} \\
  
  \hline
  
Phase Noise &
  -95 dBc/Hz &
  Ensures readout noise limited by cold amplifier \\ \hline
\begin{tabular}[c]{@{}l@{}}Recover \\ Amplitude \\ and Phase\end{tabular} &
  I/Q Channels &
  \begin{tabular}[c]{@{}l@{}}Determine frequency shift from \\ photon event using\\ phase and amplitude\end{tabular} \\ \hline
Frame Rate &
  10 kHz &
  Rate to fully process $2^{19}$ Samples \\ \hline
\end{tabular}
\label{tab1}
}
\end{table}

\section{Development of Digital Signal Processing Algorithms on FPGAs for Spaceflight}

We are developing hardware description language (HDL) algorithms for the Habitable Worlds Observatory (HWO) and the Origins Space Telescope, based on the system architecture and design parameters outlined by PRIMA~\cite{leisawitz2019origins, leisawitz2021origins}. These science measurement algorithms are implemented on field-programmable gate arrays (FPGAs) and are specifically optimized for the signal characteristics and readout requirements of MKIDs~\cite{bradley2021advancements}.

The primary objective of MKID readout is to monitor the complex transmission (or reflection) of a superconducting resonator, whose frequency shifts in response to photon absorption. Each resonator has a unique resonant frequency; photon-induced breaking of Cooper pairs alters the kinetic inductance, shifting this frequency. A probe tone near resonance is transmitted, and the output is demodulated into in-phase ($I$) and quadrature ($Q$) components, yielding a complex signal in the $I$-$Q$ plane. The resonator phase is computed as

\begin{equation}
\phi(t) = \arctan\left(\frac{Q(t)}{I(t)}\right),
\end{equation}

where $\phi(t)$ reflects shifts in resonant frequency due to absorbed photon energy. Continuous monitoring of $I(t)$ and $Q(t)$ enables detection and characterization of photon events. Both phase and amplitude variations provide energy and timing information described in \cite{day2003broadband}.

\begin{figure}[h]
\centering
\includegraphics[width=2.5cm]{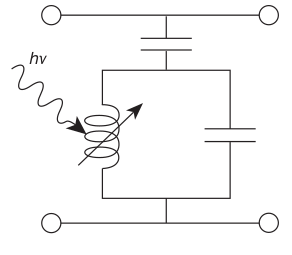}
\includegraphics[width=4.5cm]{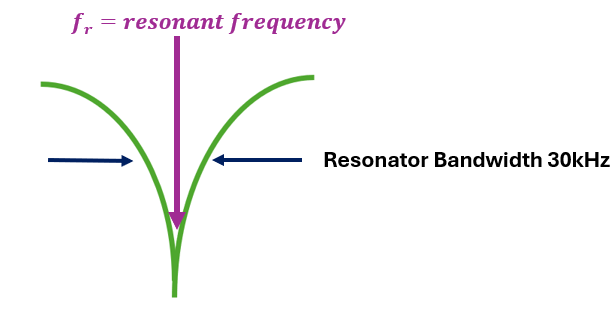}

\caption{\textbf{Single Resonator Tone}: Left: Schematic view of Kinetic Inductance Detector (from \cite{day2003broadband}): Photons with energies greater than twice the gap energy of the superconducting resonator film ($f_0$) that is cooled below its critical temperature ($T<<T_c$), are absorbed and break the Cooper pairs. The resulting increase in the quasiparticle density, changes the surface impedance of the film in the microwave resonant circuit (shown here as a parallel LC circuit capacitively coupled to the through line). The resonator bandwidth is 30 kHz. According to Table \ref{tab1}, the MKID Precision Tone Placement and Recovery Requirement is one-third of the MKID bandwidth, which corresponds to 10 kHz. This requirement directly drives the system readout specifications.}
\label{figure:SingleKid}
\end{figure}

Due to the uniformity of the MKID architecture and signal characteristics, defining the precision requirements for measuring a single MKID effectively establishes the requirements for all MKIDs in the array. As shown in Figure \ref{figure:SingleKid}, a single resonator has a bandwidth of 30 kHz. The tone precision placement requirement—which specifies the frequency resolution needed to accurately detect each distinct resonator tone—is set to one-third of the resonator bandwidth, or $\Delta f = 10 \, \text{kHz}$.

According to PRIMA specifications, a single readout chain supports up to 1,008 science resonators, with an additional 392 tones reserved for blind and calibration purposes, totaling 1,400 tones (Table~\ref{tab:PRIMAReq}) across a 2.5 GHz bandwidth. To satisfy the Nyquist criterion and ensure accurate digitization without aliasing, the system must operate with a minimum sampling frequency of $F_s$ = 5 GHz.

This resolution requirement ($\Delta f = 10 \, \text{kHz}$) and the Sampling Frequency ($F_{s}$ = 5\,GHz ) inform the necessary observation window length, or equivalently the number of samples required for tone recovery via FFT-based analysis:

\begin{equation}
    N = \frac{F_s}{\Delta f} = \frac{5 \times 10^9}{10 \times 10^3} = 2^{19} \, \text{samples}.
\end{equation}

Hence, each science measurement must process at least \(2^{19}\) samples to achieve the desired frequency discrimination. This defines the baseline frequency resolution and temporal window required for MKID tone detection and spectral analysis across the full MKID array.

Figure~\ref{figure:SysBlkMKIDUp} depicts the FPGA-based system block diagram for processing two data streams in the MKID readout, designed to capture and process the measurement signals. Algorithms highlighted in blue (Pulse Detection and Tone-tracking) will be implemented in HWO, not PRIMA. A compilation of resonator drive tones in the time domain are synthesized for the MKID array using a combination of a CORDIC (COordinate Rotation DIgital Computer) algorithm and a high-performance synthesis Fast Fourier Transform (FFT) implementation, structured as a polyphase filter bank (PFB). The resulting complex $I/Q$ waveform is digitally generated and transmitted to the MKID array via two 14-bit digital-to-analog converters (DACs), specifically the \texttt{TI-DAC12DL3200}.

The returning complex $I/Q$ analog signal, spanning a frequency range from 2.4 GHz to 4.9 GHz, is sampled using two dual-channel 12-bit analog-to-digital converters (ADCs), the \texttt{TI-ADC12DJ5200}, which are optimized for wideband digitization and direct RF sampling.

\begin{figure}[h]
\centering
\includegraphics[width=9cm]{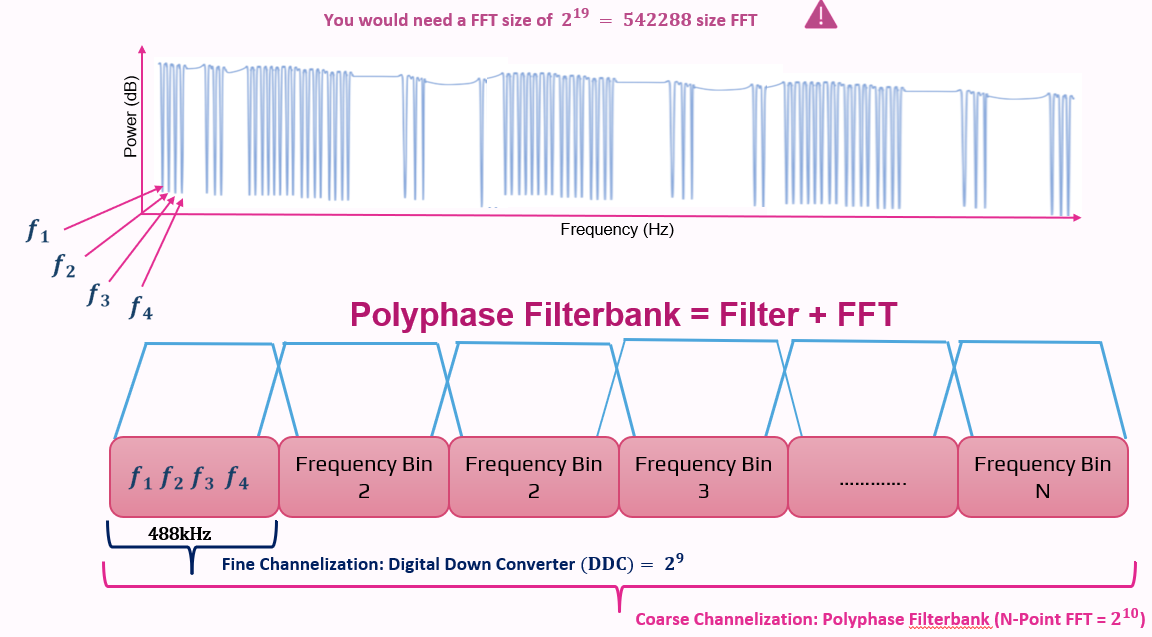}
\caption{A two-stage channelization process is used to efficiently achieve the required 10~kHz spectral resolution for MKID readout. The measurement is based on \(2^{19}\) samples with a sampling rate of \(F_s = 5\,\text{GHz}\), yielding a resolution of \(\frac{F_s}{2^{19}} = 10\,\text{kHz}\). The first stage employs a \(2^{10}\)-point polyphase filterbank (PFB) for coarse channelization, where MKIDs may share bins. The second stage performs fine channelization using a \(2^9\)-point FFT per bin to resolve individual tones. This hierarchical approach is significantly more resource-efficient than a single \(2^{19}\)-point FFT, which exceeds typical FPGA capabilities.}

\label{figure:coarse/fine}
\end{figure}

The incoming complex signal is sampled at a rate of \(F_{s} = 5\,\mathrm{GHz}\). Frequency decomposition is performed on \(2^{19}\) samples in two stages (Figure \ref{figure:coarse/fine}), rather than using a single very long FFT, which would consume excessive FPGA resources. This two-tiered spectral decomposition approach enables high dynamic range and scalable readout of thousands of frequency-multiplexed MKID channels within the constraints of FPGA-based hardware.

\subsubsection{Coarse and Fine Channelization}

\begin{figure}[h]
\centering
\includegraphics[width=8.5cm]{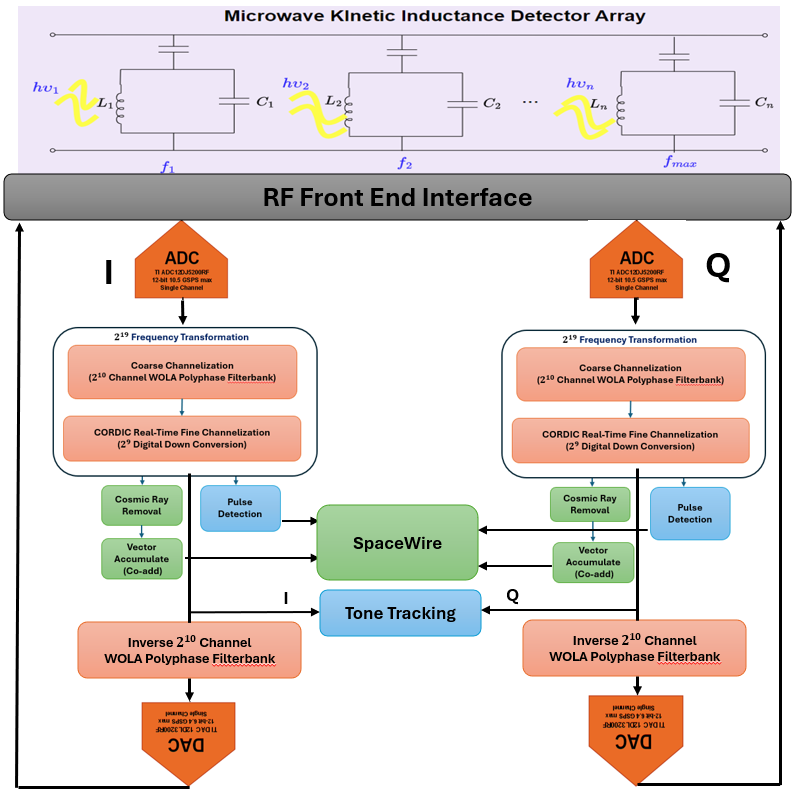}
\caption{\textbf{Space-Qualified FPGA-Based Dual-Chain MKID Readout System:} Data is continuously transmitted and received via space-qualified DACs and ADCs, respectively, with a sampling frequency of \(F_{s} = 5\,\mathrm{GHz}\), and is processed in chunks of \(2^{19}\) samples in accordance with the requirements outlined in Table~\ref{tab:PRIMAReq}. The transmitted \(I/Q\) signals consist of a time-domain compilation of frequency tones generated by the CORDIC and the inverse Polyphase Filterbank, which are output through the DAC. The received \(I/Q\) signals are digitized via the space-qualified ADCs and then undergo a \(2^{19}\)-point frequency transformation performed in two stages: coarse and fine channelization. A cosmic ray removal algorithm is applied at each phase increment of the fine channelization frame, prior to the vector accumulate (co-add) step.
Both the I and Q channel data are inputs to the tone tracking algorithm. The Co-add and pulse detection data are transmitted via SpaceWire.}
\label{figure:SysBlkMKIDUp}
\end{figure}

Coarse channelization is implemented using an oversampled polyphase filterbank (PFB) with \(2^{10}\) channels. A polyphase filterbank applies a filter around each FFT frequency bin to mitigate spectral leakage. Multiple MKIDs share FFT bins at this stage. To resolve individual resonator frequencies within each FFT bin, a digital downconversion (DDC) algorithm with \(2^{9}\) points is implemented using a CORDIC processor. Specifically, \(2^{9}\) frames of the \(2^{10}\) channel PFB output are used in the fine channelization (DDC) algorithm to extract each frequency component from the shared FFT bins. This step, enables the precise recovery of individual MKID tones in the I/Q channels, allowing for accurate measurement of amplitude shifts associated with incident photons. During this DDC stage, vector accumulate (co-add) is also performed to improve the signal-to-noise ratio, enhancing the fidelity of photon detection.
Additionally, cosmic ray mitigation algorithms operate on each MKID resonator during the DDC process. A continuous processing window of $2^{19}$ samples enables simultaneous removal of cosmic ray events and extraction of both power changes---via coarse and fine channelization---and phase changes---via I/Q channels. This window length supports data transmission over SpaceWire at a 10\,kHz frame rate  as illustrated in Figure \ref{figure:SysBlkMKIDUp}.

\subsubsection{Cosmic Ray Removal}

Cosmic ray interactions with MKID arrays produce abrupt, transient disturbances, typically observed as sudden phase shifts or amplitude changes in individual resonators. These glitches can result from energy deposition in the substrate, often causing simultaneous disturbances across multiple neighboring MKIDs. To mitigate their impact, cosmic ray glitch removal will be performed in real time onboard the FPGA using the high-rate data stream prior to the vector accumulate (co-add) depicted in (Figure \ref{figure:SysBlkMKIDUp}). An FPGA-based algorithm is currently under development that uses matched filtering with a modeled glitch template to cross-correlate against the incoming data stream, enabling real-time detection—even for low-amplitude events. Future enhancements may include the exploration of more advanced techniques beyond basic matched filtering.

\subsubsection{Pulse Detection and Tone-Tracking for HWO}
A pulse detection algorithm is executed at each phase increment of the fine channelization to capture the energy profile of each detector. In parallel, real-time phase information from the I/Q channels is used for tone tracking, enabling accurate determination of frequency shifts in each resonator and precise photon event detection. While these algorithms are not currently baselined for PRIMA, they are planned for future implementation on HWO.

\section{Latest Results}

\begin{figure}[h]
\centering
\includegraphics[width=9cm]{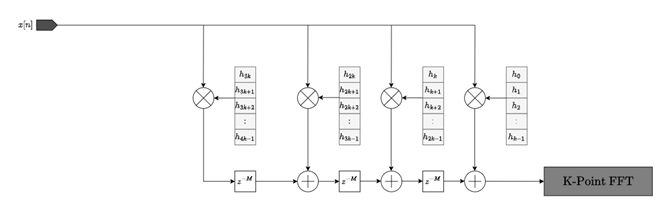}
\caption{The hardware implementation of the critically sampled polyphase filterbank \cite{harris2022multirate}  uses a ROM to store filter coefficients arranged in a polyphase structure. During each clock cycle, the system cycles through the appropriate coefficient phases, enabling efficient real-time processing with only four multiplication and addition operations. This architecture significantly reduces resource utilization while maintaining full-rate throughput.}
\label{fig:fpga_cs}
\end{figure}

\begin{figure}[h]
\centering
\includegraphics[width=9cm]{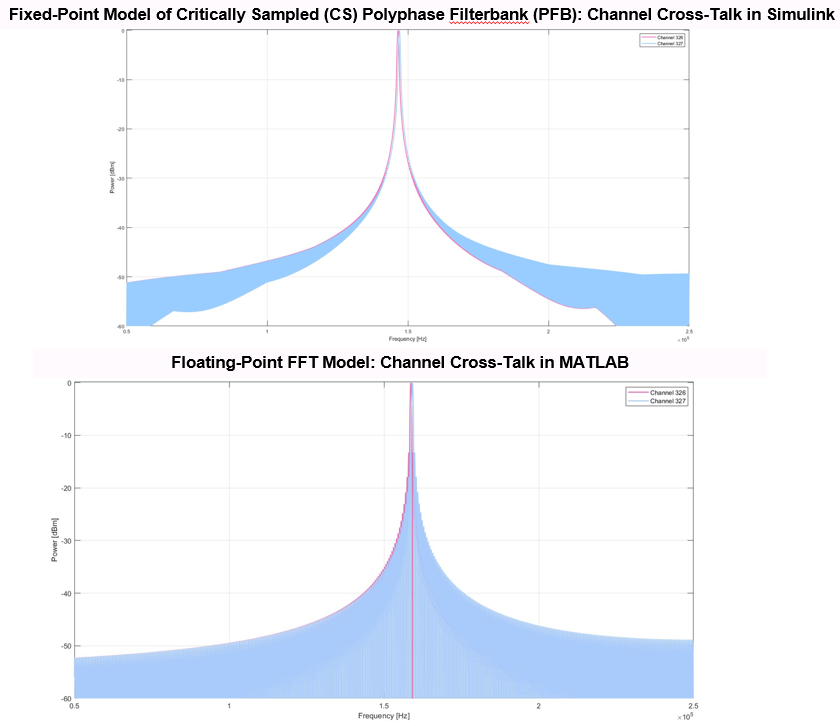}
\caption{Cross-talk response between two adjacent channels in the Fixed-Point Simulink model of the Critically Sampled Polyphase Filterbank (top) versus the Floating-Point FFT Model (bottom). The measured cross-talk level is approximately \(-48\,\text{dBc}\).}
\label{fig:cross_talk}
\end{figure}

The critically sampled polyphase filterbank (CS-PFB)—a necessary precursor to the final oversampled design—was successfully implemented in hardware using fixed-point arithmetic and a resource-efficient architecture. As shown in Figure~\ref{fig:fpga_cs}, the design utilizes a ROM to store filter coefficients in polyphase format. During each clock cycle, the system cycles through these coefficients, performing real-time convolution using only four multipliers and adders. This approach significantly reduces logic utilization while sustaining full-throughput operation at the system's input rate, making it well-suited for FPGA-based deployment in constrained environments such as space-based MKID readout systems. To evaluate channel isolation, the cross-talk response between two adjacent channels was simulated in a fixed-point MATLAB Simulink model of the CS-PFB. The results, shown in Figure~\ref{fig:cross_talk}, indicate a cross-talk suppression level of approximately \(-48\,\text{dBc}\). This level of isolation confirms that the fixed-point implementation maintains sufficient spectral separation to meet the MKID tone recovery requirement of 10~kHz, even in the presence of quantization and arithmetic precision constraints. These results validate both the architectural efficiency and signal integrity of the CS-PFB design and demonstrate its viability for high-density, frequency-multiplexed detector readout on FPGA platforms.

\bibliographystyle{IEEEtran}
\bibliography{report}

\end{document}